# Proposed Video Encryption Algorithm v/s Other Existing Algorithms: A Comparative Study

Ajay Kulkarni
Vidyalankar Inst. Of Tech.
Mumbai, India

Saurabh Kulkarni
VES Inst. Of Tech.
Mumbai, India

Ketki Haridas
Vidyalankar Inst. Of Tech.
Mumbai, India

Aniket More
Vidyalankar Inst. Of Tech.
Mumbai, India

## ABSTRACT
Securing multimedia data has become of utmost importance especially in the applications related to military purposes. With the rise in development in computer and internet technology, multimedia data has become the most convenient method for military training. An innovative encryption algorithm for videos compressed using H.264 was proposed to safely exchange highly confidential videos. To maintain a balance between security and computational time, the proposed algorithm shuffles the video frames along with the audio, and then AES is used to selectively encrypt the sensitive video codewords. Using this approach unauthorized viewing of the video file can be prevented and hence this algorithm provides a high level of security. A comparative study of the proposed algorithm with other existing algorithms has been put forward in this paper to prove the effectiveness of the proposed algorithm.

## 1. INTRODUCTION
With the fast growth of multimedia technology many armies across the world are using videos to train newly recruited troops. Such sensitive data has to be protected either in transmission or storage. One possible way to protect multimedia information is to stop unauthorized access. But this approach cannot make sure that the multimedia information is physically secure. Another easy approach is to encrypt the complete bit stream with a cryptographic algorithm, such as DES or AES. However videos generally possess a large amount of data and require real-time operations. Moreover, in the case of the wireless mobile systems, there is limited processing power, memory and bandwidth, and is rarely able to handle the heavy encryption processing load. Therefore, taking into consideration the specific characteristics for resource-limited systems, new video encryption algorithms need to be developed. For real-world applications, a video encryption algorithm has to take into account various parameters like security, computational efficiency, compression efficiency and so on. Different types of video applications require different levels of security. For example, for Video on Demand, low security will be fine, whereas for military purposes or financial information, high level of security is required to completely prevent unauthorized access. Computational efficiency means that the encryption or decryption process should not cause too much time delay, so that the requirements of real-time applications are met. Video compression is employed to reduce the storage space and save bandwidth, so that the encryption process should have the least impact on the compression efficiency. All in all, a well-designed video encryption algorithm should provide sufficient security, high computational efficiency; impose little impact on the compression efficiency. In this paper, the working of the proposed algorithm and how it is better than the existing algorithms has been explained.

## 2. CLASSIFICATION
Video encryption algorithms can be classified into four basic categories:

### 2.1 Completely Layered Encryption
In this method, the entire video is first compressed and is then encrypted using traditional algorithms like RSA, DES, and AES. This technique is not applicable in real time video applications due to heavy computation and very low speed.

### 2.2 Encryption Using Permutation
Here, the video content is scrambled using a permutation algorithm. The entire video content may be scrambled or only particular bytes. A permutation list maybe used as a secret key for encryption.

### 2.3 Selective Encryption
To save computational complexity only particular video bytes maybe encrypted.

### 2.4 Perceptual Encryption
After encryption using this technique the video will still be perceptible. The audio/video quality can be controlled continuously.

## 3. PERFORMANCE PARAMETERS
To evaluate and compare video encryption algorithms there is a need to define a set of performance parameters.

### 3.1 Encryption Ratio
ER gives the ratio of size of encrypted video to the size of the original video. Lesser the ER better is the computational efficiency of the algorithm.

### 3.2 Compression Efficiency
The ease of compression depends on the data compression efficiency. Some encryption algorithms introduce additional information that is necessary for encryption/decryption. The size of the encrypted video should be as less as possible.

### 3.3 Degradation
This criterion measures the distortion of the video with respect to the original video. Visual degradation should be achieved to a considerable level so that the video is not understandable to the attacker. In highly confidential videos, high visual degradation is a must.

### 3.4 Security
The algorithm should be resistant to attacks such as brute-force and known-plaintext attack.

### 3.5 Format Compliance
Encrypted bit stream must be compliant with the compressor. The standard decoder should be able to decode the encrypted videos.

### 3.6 Speed
For real-time applications the encryption and decryption time should be as less as possible.

## 4. PROPOSED ALGORITHM
In the encryption block the video is first divided into frames by using a video cutter. These frames also contain the audio information. Here the individual frames are in video format instead of being an image because image files cannot store audio information. The shuffling block then shuffles these frames and then these frames are passed on to the frame stitching block. These frames that are in random position form the new video. By doing this, an audio stream that is impossible to decrypt and understand is achieved.

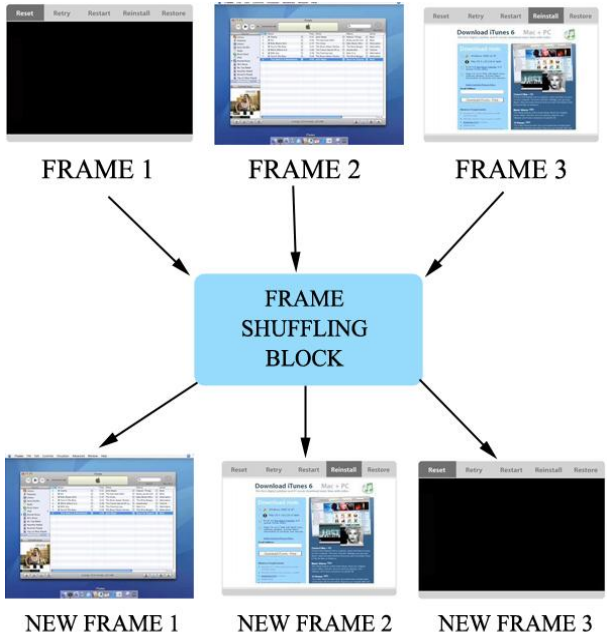

**Fig 1: Frame shuffling block**

Thus it is clear from Fig. 1 and 2 that the audio is very difficult to decode unless the shuffling methodology is known. The shuffling algorithm uses a random key generation function implemented in java; this function is termed as the "Shuffling Key" and is sent along with the video to the destination decryption block. The shuffling key is encrypted along with the video using AES. Hence by doing this it is ensured that the audio is impossible to understand and the video just shows random frames. Brute-force attacks have become more sophisticated, groups of expert video analysts may sit together and analyse the entire video frame by frame and may bring together the original video. Hence there is a need to increase the security further to prevent such Brute force attacks. To do this AES is required. AES is used to encrypt the codewords extracted from MVDs, DCs and ACs. Codewords are a stream of digital bits. These digital bits contain all the information of an image. The AES algorithm will be used along with the individual encoding algorithms of ACs, DCs and MVDs. Only the important or sensitive codewords are extracted and encrypted so that computation time is saved. After encryption, the codewords are jumbled up and the video will like layers of scattered colours. The blocks on each frame will be the same but their position will be changed. Thus the video is beyond human interception.

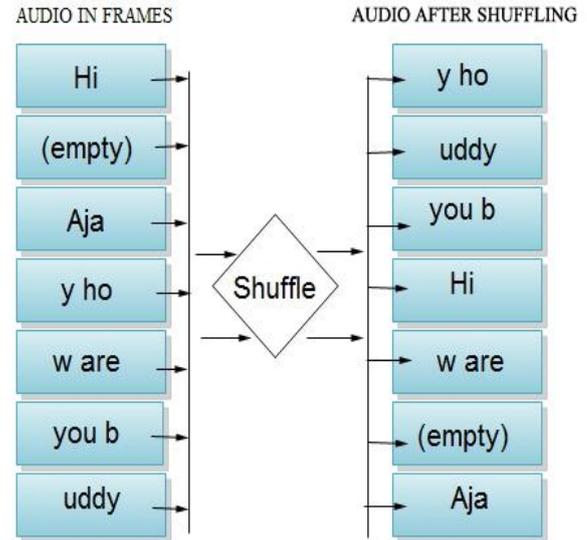

**Fig 2: Audio after frame shuffling**

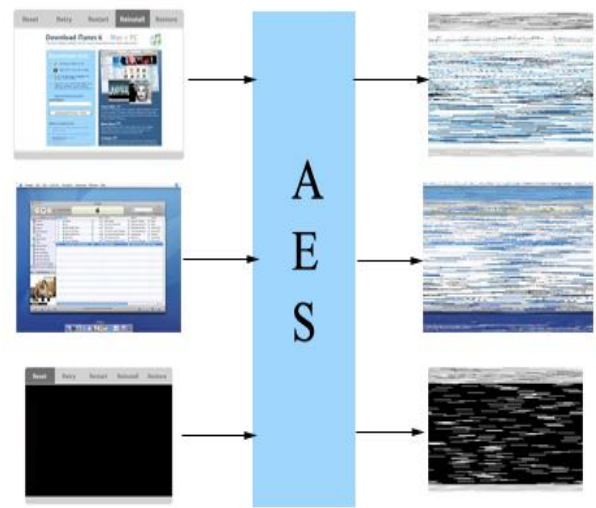

**Fig 3: Frames after AES Encryption**

Finally after the video is transferred to the client, the decoder will first run the AES algorithm over the codewords to decode and obtain the clean video. The decryption block will also decode the random key and use it to reshuffle the frames to its original position. The frame stitching block once again comes into existence at the decoding side to stitch the frames and output the original video. The complete process of encryption and decryption is explained in Fig 4.

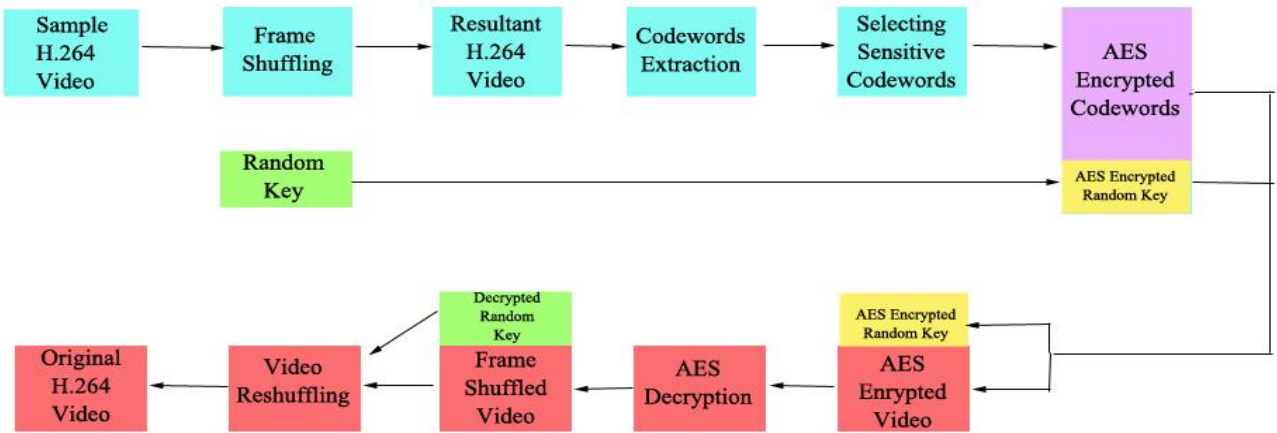

**Fig 4: Block Diagram of proposed algorithm**

## 5. ANALYSIS AND RESULTS

To evaluate the performance of this algorithm a sample .mp4 video, 108 seconds long, which was encoded using H.264 was used. Using MediaCoder the video is split into smaller videos containing one frame each. Using random key generation function in Java the video files are shuffled. Using PhotoLapse all the videos are stitched together to make a video with shuffled frames. The codewords of the resultant video was then completely encrypted using VirtualDub. The table below shows the time in ms spent on each stage. All these computations were done on a standard computer.

**Table 1. Computation time of each task during Encryption**

| Task | Computation Time (ms) |
| --- | --- |
| Video Shredding | 1210 |
| Shuffling | 113 |
| Video Stitching | 1096 |
| AES Encryption | 1580 |

The total time required for encryption of a 108 second video was found to be 3999 ms which is approximately 4 seconds. That means approximately 3.7% of the video length is required to encrypt the video. Decryption takes exactly the same amount of time as it is just the reverse process. All in all it took just took around 8 seconds to encrypt and decrypt a 108 second video. Thus this algorithm is efficient enough when it is used in new generation smart phones. Although the encryption time is slightly higher, the type of data protection it provides is unmatched.

## 6. EXISTING ALGORITHMS

Various algorithms have been proposed in the near future. Some look very effective but lack efficiency. Listed below are a few existing algorithms.

### 6.1 Simple Permutation

This proposed method [1][2] encrypts every byte in the video stream using algorithms such as AES or DES. This algorithm considers the video bit stream as standard text data. The security level is high as every byte is encrypted one by one. Encryption algorithms such as AES and DES are break proof. This algorithm is not practical as encrypting large videos will take a very long time. Simple Permutation is not suitable for real-time applications as the time factor is very important. As the video stream is encrypted after compression there is zero effect on compression efficiency.

### 6.2 Pure Scrambling

Video bytes in each frame of the video are shuffled using permutation operation. This proposed method [3] is very handy in applications where hardware decodes the video. But in day to day application decryption is the work of software. This method is susceptible to the known-plaintext attack and hence should be used with caution. The permutation sequence can be easily figured out by comparing the known frames with the cipher text. After understanding the sequence, the attacker can easily decrypt the entire video.

### 6.3 Crisscross Permutation

This proposed algorithm [4] first generates a 64 byte permutation list. This list is then quantized into an 8x8 block. This is followed by a simple splitting procedure. The random permutation list is then applied to the split blocks and the result is then encoded. Computational complexity is relatively low and hence the encryption and decryption process is not too complex. Crisscross permutation distorts the DCT coefficients and hence the video compression rate is lowered. This algorithm also cannot withstand the known-plaintext attack.

### 6.4 Choose and Encrypt

Encrypting and decrypting the entire video stream is not practical in real-time applications. A solution [5] is needed in which frames in the video can be selectively encrypted. By implementing such a methodology the complexity and encryption/decryption overhead is decreased to a great level. However, the level of security should also be maintained. This algorithm can be successful if a proper tradeoff can be maintained between complexity and security.

### 6.5 Other Proposed Algorithms

Many authors and research scholars have proposed video encryption algorithms in the recent past. Let us have a look at some of these innovative algorithms and understand the working of each one of them.

### 6.5.1 *Methodology proposed by Bergeron and Lamy-Bergot*

A syntax based encryption algorithm is proposed for H.264 videos [6]. Encryption is done in the encoder. The proposed method inserts the encryption mechanism within the video encoder, providing secure transmission which does not hamper the transmission process. The bits selected for encryption are chosen with respect to the considered video standard according to the following rule: each of the encrypted configurations gives a synchronized and a standard compliant bit stream. This can in particular be done by encrypting only parts of the bit stream which have no or a negligible impact in evolution of the decoding process, and whose impact is consequently purely a visual one.

### 6.5.2 *Methodology proposed by Lian, Liu, Ren and Wang*

This scheme is proposed for AVC [7]. During AVC encoding sensitive data such as intra prediction mode, residue data and motion vector are encrypted partially. DCs are encrypted based on context based adaptive variable length coding. The encryption scheme is of high key sensitivity, which means that slight difference in the key causes great differences in encrypted video and makes statistical attack difficult. It is difficult to apply known plaintext attack. In this encryption scheme, each slice is encrypted under the control of a 128 bit sub-key. Thus, for each slice, the brute force space is $2^{128}$. This brute force space is too large for attackers to break the cryptosystem. According to the encryption scheme proposed here, both the texture information and the motion information are encrypted, which make it difficult to recognize the texture and motion information in the video frames.

### 6.5.3 *Methodology proposed by Lian,, Sung and Wang*

This scheme is proposed for 3D-SPIHT videos [8-10]. In this scheme different number of wavelet coefficients encrypts different number of coefficients signs and data cubes. Videos can be degraded to different degrees under the control of quality factor. Its encryption strength can be adjusted according to certain quality factor. It is not secure against known chosen plaintext attack.

### 6.5.4 *Methodology proposed by Li, Chen, Cheung, Bharat Bhargava, and Kwok-Tung Lo*

This design is a generalized version for perceptual encryption, by selectively encrypting FLC data elements in the video stream [11]. Apparently, encrypting FLC data elements is the most natural and perhaps the simplest way to maintain all needed features, especially those needed for strict size preservation. To maintain format compliance, only last four FLC data elements are considered, which are divided into three categories i) intra DC coefficient ii) sign bits of non intra DC coefficients and AC coefficients iii) sign bits and residuals of motion vectors.

## 7. COMPARISON
Refer Table 2 & 3.

## 8. CONCLUSION
In this paper the proposed algorithm was compared with the currently known methods of cryptography. The two different types of the encryption methods (Symmetric key encryption and Asymmetric key encryption) were highlighted and evaluated with respect to their security level and encryption speed. Also, various currently existing algorithms have been explained. From the table it is evident that Simple Permutation algorithm [1][2] and the proposed video encryption algorithm are the most secure algorithms, whereas crisscross permutation algorithm [4] has a serious security flaw; it is not immune to the known-plaintext-attack. With respect to encryption speed, the proposed encryption algorithm and crisscross permutation algorithm [4] are fast, Simple Permutation [1][2] is very slow while applying DES on entire video stream. Summarizing, a trade-off needs to be maintained in video encryption algorithms and its choice depends on the applications. But for military applications, the proposed algorithm will be most suitable as it provides high level of security with a good computation speed.

**Table 2. Comparisons of Video Encryption Algorithms**

| METHODOLOGY | SECURITY LEVEL | SPEED | VIDEO SIZE | ECRYPTION RATIO |
|---|---|---|---|---|
| SIMPLE [1][2] | HIGH | SLOW | NO CHANGE | 100% |
| PURE [3] | LOW | FAST | NO CHANGE | 100% |
| CRISSCROSS [4] | VERY LOW | VERY FAST | BIG CHANGE | 100% |
| CHOOSE & ENCRYPT [5] | HIGH | FAST | NO CHANGE | 50% |
| PROPOSED ALGORITHM | VERY HIGH | FAST | NO CHANGE | 100% |

**Table 3. Comparison of Encryption Algorithms**

| ALGORITHM | COMPLEXITY | SPEED | MEMORY | KEY TYPE | KEY LENGTH | SECURITY LEVEL |
|---|---|---|---|---|---|---|
| DES | Complex | High | N/A | Private | 56 bit | Low |
| AES | Complex | High | Very Low | Private | 128, 192, 256 bit | High |
| RSA | Simple | High | N/A | Public | Variable | High |